\documentclass[aps,prl,reprint,groupedaddress,amsmath]{revtex4-1}

\usepackage{graphicx}

\usepackage{layouts}

\begin{document}


\title{Sensing of Static Forces with Free-Falling Nanoparticles}



\author{Erik Hebestreit}
\affiliation{Photonics Laboratory, ETH Z\"{u}rich, 8093 Z\"{u}rich, Switzerland}

\author{Martin Frimmer}
\affiliation{Photonics Laboratory, ETH Z\"{u}rich, 8093 Z\"{u}rich, Switzerland}

\author{Ren\'{e} Reimann}
\affiliation{Photonics Laboratory, ETH Z\"{u}rich, 8093 Z\"{u}rich, Switzerland}

\author{Lukas Novotny}
\homepage[]{http://www.photonics.ethz.ch}
\affiliation{Photonics Laboratory, ETH Z\"{u}rich, 8093 Z\"{u}rich, Switzerland}


\date{\today}

\begin{abstract}
Miniaturized mechanical resonators have proven to be excellent force sensors. However, they usually rely on resonant sensing schemes, and their excellent performance cannot be utilized for the detection of static forces. Here, we report on a novel static-force sensing scheme and demonstrate it using optically levitated nanoparticles in vacuum. Our technique relies on an off-resonant interaction of the particle with a weak static force, and a resonant read-out of the displacement caused by this interaction. We demonstrate a force sensitivity of $10\,\mathrm{aN}$ to static gravitational and electric forces acting on the particle. Our work not only provides a tool for the closer investigation of short-range forces, but also marks an important step towards the realization of matter-wave interferometry with macroscopic objects.
\end{abstract}


\maketitle



Despite our solid understanding of physics at macroscopic scales, interactions between microscopic objects at short distances still bear countless secrets \cite{French2010,Klimchitskaya2009,Volokitin2007}. Micro- and nanomechanical sensors provide astounding force sensitivities, which aim to ultimately reveal these secrets \cite{Mamin2001,Stipe2001a,Li2007,Gavartin2012,Moser2013,Norte2016}. Amongst these sensors are optically trapped dielectric nanoparticles in vacuum. The motion of these particles can be optically measured and controlled with remarkable precision \cite{Li2011,Gieseler2012}. Furthermore, due to the absence of mechanical clamping losses, the particle resembles a high-quality mechanical resonator, rendering it ideally suited for force-sensing applications. Levitated particle sensors have demonstrated zepto-Newton sensitivity \cite{Ranjit2016} and have been suggested for numerous high-precision experiments, ranging all the way from the detection of Casimir or van der Waals interactions \cite{Geraci2010}, and non-Newtonian forces \cite{Geraci2015}, to the production and sensing of mechanical quantum states in macroscopic objects \cite{Chang2010,Romero-Isart2011,Scala2013,Bateman2014}.

Most force sensors, including levitated particles, rely on a resonant sensing scheme. Here, the sensor's intrinsic resonance is harnessed to strongly amplify the response to a perturbation \cite{Moser2013,Ranjit2016}. This scheme implies a trade-off between the sensor's measurement bandwidth and its sensitivity. While this technique allows great sensitivity to forces oscillating close to the resonance frequency of the resonator, weak forces at low frequencies are difficult to detect. In particular, resonant sensing fails for truly static forces. As recently demonstrated, their detection requires to measure the minute displacement of the sensor as a response to the force \cite{Blums2017}. Yet, it is an open question how the performance of resonant sensing schemes can be transferred to the static case.

In this Letter, we propose and demonstrate a force sensing scheme, that transfers the superior performance of resonant sensors to the realm of static interactions. By temporarily reducing the sensor's resonance frequency to zero, we allow the sensor to freely interact with the static force. When subsequently restoring the spring constant to its original value, we are able to resonantly read out the sensor displacement with a high precision. We implement this novel static-force sensing technique using an optically levitated nanoparticle, whose spring constant can be modulated in a wide range by tuning the intensity of the laser beam used for trapping the particle. Such a modulation of the trapping potential has already shown to be a powerful tool for measuring the energy of atom clouds and even single atoms using release-recapture thermometry \cite{Chu1985,Mudrich2002}. Combining this method with the precise position measurement of levitated particles, we achieve a static-force sensitivity of $10\,\mathrm{aN}$. With further refinements our sensing scheme should be able to sense static forces in the zepto-Newton regime. We demonstrate our sensor performance by detecting the gravitational interaction between the levitated particle and the earth, as well as the Coulomb force acting on a charged particle in an electric field.

\begin{figure} 
	\includegraphics{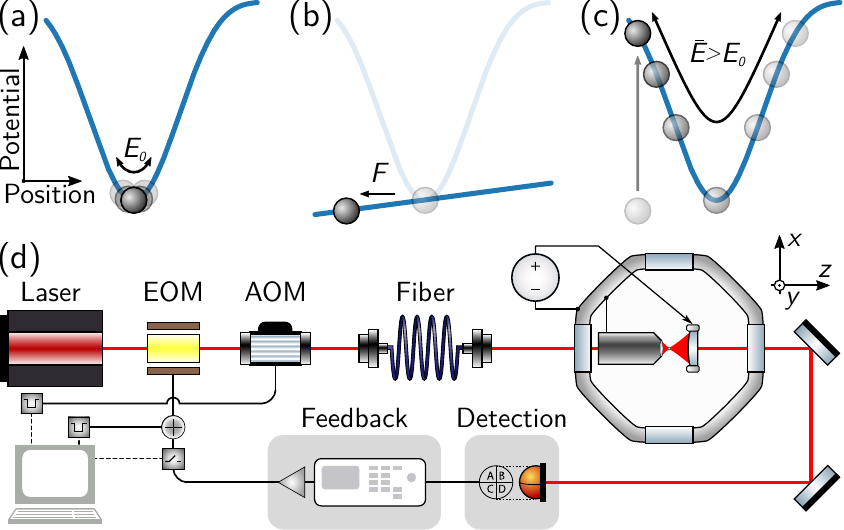}
	\caption{(a)--(c)~Scheme for static-force sensing. (a)~To initialize the system, we trap a mass (nanoparticle) with low energy $E_0$ in a harmonic potential (blue). (b)~We deactivate the trapping potential for a duration $\tau$, during which the particle only feels the force $F$ that is to be measured, and gets displaced. (c)~When reactivating the optical trap, the particle gains potential energy due to its displacement from the trap center. As a result, the particle oscillates with a higher amplitude than in (a), i.e., with an increased center-of-mass (COM) energy $E$. (d)~Experimental setup. The nanoparticle is levitated in the focus of a laser beam inside a vacuum chamber. We switch the optical trap off with an electro- and acousto-optical modulator (EOM, AOM). To measure the particle position, we detect the scattered light from the particle with a balanced detector. A parametric feedback uses the position information to reduce the COM energy by modulating the trapping beam. Applying a voltage between the objective and the holder of the collection lens allows us to create a static electric field that can be used to exert a static force on a charged particle. \label{fig:setup}}
\end{figure}

The principle of our force-sensing scheme is illustrated in Fig.~\ref{fig:setup}(a). We prepare a mass in a harmonic potential with a low oscillation energy $E_0$. The potential is stiff enough to render the displacement of the mass due to the static force (which is to be measured) negligible.  Then, we turn off the harmonic potential for an interaction time $\tau$ [see Fig.~\ref{fig:setup}(b)]. The force
$F$ causes a displacement of the mass, which upon reactivating the harmonic potential results in an oscillation at the resonance frequency with an increased amplitude compared to the initial state, as illustrated in Fig.~\ref{fig:setup}(c). In the case of weak damping, we can measure this amplitude with high precision to deduce the magnitude of the static force.


\paragraph{Experimental setup.} 
We demonstrate our static-force sensing scheme using an optically levitated silica nanosphere with a measured mass of $m=2.0(7)\,\mathrm{fg}$ in vacuum \cite{Li2011,Gieseler2012,Ranjit2016}. A schematic of the experimental setup is shown in Fig.~\ref{fig:setup}(d). The three-dimensional oscillator potential is formed by a laser beam ($140\,\mathrm{mW}$, linear polarization) that is strongly focused (numerical aperture $0.85$) and results in oscillation frequencies of $\Omega_{z,0}=2\pi\times60\,\mathrm{kHz}$ along the optical axis, and $\Omega_{x,0}=2\pi\times195\,\mathrm{kHz}$ and $\Omega_{y,0}=2\pi\times160\,\mathrm{kHz}$ in transverse directions. In this trapping potential, a static force of $10\,\mathrm{aN}$ causes a particle displacement of $5\,\mathrm{pm}$, which is small compared to any oscillation amplitudes we encounter in this work. Therefore, we can neglect the influence of the static force whenever the trapping potential is activated. Using intensity modulators, we can reduce the trapping power to below $100\,\mathrm{nW}$ and therewith the optical forces to less than $0.2\,\mathrm{aN}$, which is more than 100 times weaker than the gravitational force between the particle and the earth. To minimize interactions of the particle with the surrounding gas, we lower the gas pressure to below $10^{-5}\,\mathrm{mbar}$. Collecting the light scattered from the particle with a balanced detector provides us with the particle's center-of-mass (COM) position \cite{Gieseler2012}. We use this position information to generate a feedback that cools the particle's COM motion to less than $100\,\mathrm{mK}$ \cite{Jain2016}. In order to minimize possible electrostatic interactions with the environment, we discharge the nanoparticle prior to our experiment \cite{Frimmer2017}.

\paragraph{Gravitational force measurement.} 

\begin{figure*} 
	\includegraphics{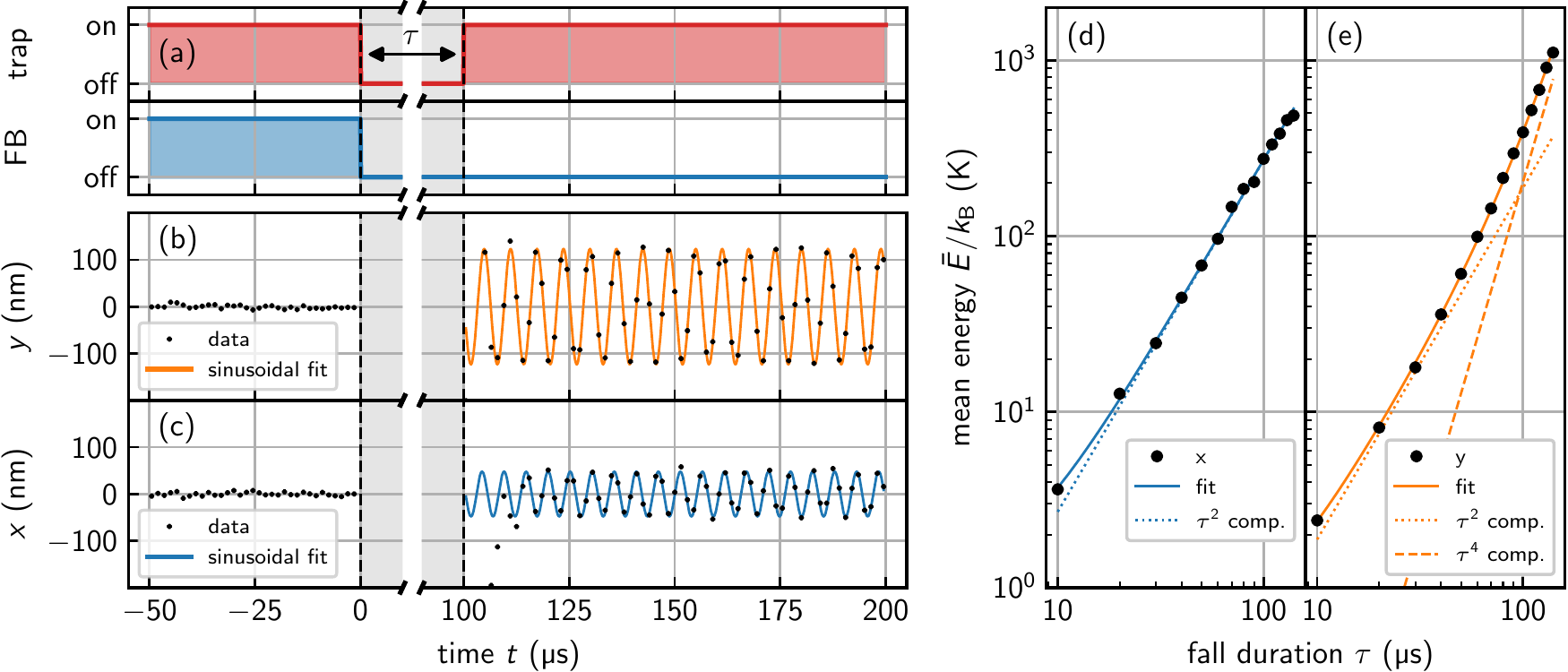}
	\caption{(a)~Timing scheme of a single free-fall cycle. For $t<0$, the trapping potential is on and the feedback (FB) cools the particle's COM motion [cf.~Fig.~\ref{fig:setup}(a)]. From $t=0$ to $t=\tau=100\,\mathrm{\mu s}$, we deactivate the optical trap and the feedback such that the particle can freely interact with the static force [cf.~Fig.~\ref{fig:setup}(b)]. At time $t=\tau$, the trap is reactivated. The feedback remains disabled to allow for sustained oscillations of the particle in the potential [cf.~Fig.~\ref{fig:setup}(c)]. We plot the recorded motion along the $y$ axis (b) and the $x$ axis (c) in black. For $t>\tau$, the particle undergoes sinusoidal oscillations (orange and blue fit) with a larger amplitude along the $y$ direction. (d)~Average oscillation energy $\bar{E}$ after free fall in $x$ direction for varying fall durations $\tau$ (error bars are smaller than the symbol size). The average energy increases quadratically with $\tau$ (dotted line). (e)~Along the $y$ axis, where gravity is accelerating the particle, the scaling deviates from $\tau^2$. For both axes, we fit $\bar{E}=\beta_0+\beta_2\tau^2+\beta_4\tau^4$ (solid line) and show the $\beta_2\tau^2$ term (dotted line), and the $\beta_4\tau^4$ term (dashed line). The latter only contributes for the $y$ axis measurement, and is a signature of the gravitational force acting on the particle. \label{fig:energies_free_fall}}
\end{figure*}

In a free-fall experiment, we measure the gravitational force that acts on the particle. This force amounts to $F=-mg=-20\,\mathrm{aN}$ and is oriented along the $y$ axis. Figure~\ref{fig:energies_free_fall}(a) shows the experimental protocol for a single free-fall cycle: (1)~We initialize the trapped particle to a low COM energy using the parametric feedback. (2)~At time $t=0$, we switch off both the feedback and the optical trap, to allow the particle to freely interact with the gravitational force. (3)~After the interaction time $\tau$, we retrap the particle by switching the optical trap back on while the feedback remains off. In Fig.~\ref{fig:energies_free_fall}(b), we plot the COM motion along the $y$ axis during one free-fall cycle at a pressure of $6\times10^{-6}\,\mathrm{mbar}$. For $t<0$, the COM energy is reduced to $E_{y,0}/k_\mathrm{B}=47\,\mathrm{mK}$ \cite{Hebestreit2017}. From $t=0$ to $t=\tau$, while the optical trap is deactivated, we have no information about the particle position (gray shaded area). For $t>\tau=100\,\mathrm{\mu s}$, we observe a harmonic and strongly underdamped oscillation of the particle's COM at the trap frequency $\Omega_{y,0}$ (sinusoidal fit in orange). This large oscillation amplitude in $y$ direction originates from the displacement the particle acquired due to the gravitational force [cf.~Fig.~\ref{fig:setup}(c)].

As a reference, we show in Fig.~\ref{fig:energies_free_fall}(c) the particle motion along the $x$ axis, orthogonal to the direction of gravity, where no static force is acting on the particle. Before the free fall, we measure an initial COM energy of $E_{x,0}/k_\mathrm{B}=63\,\mathrm{mK}$. After the free fall, we record an oscillation at the frequency $\Omega_{x,0}$ with an amplitude significantly larger than before the fall. At first sight, this observation seems to contradict our expectation for the behavior in the absence of a static force. Indeed, for a particle that is at rest at time $t=0$, we expect no oscillation after the free fall if no static force is acting. However, as the feedback is not perfect, the particle remains with a finite initial velocity $\dot{x}(0)$ at time $t=0$, which leads to a displacement $\Delta x=\dot{x}(0)\tau$ directly after the free fall. This displacement results in an oscillation along the $x$ direction for $t>\tau$.

Before the free fall, the feedback cooled state of the particle's COM is a thermal state, and the initial velocities $\dot{x}(0)$ and $\dot{y}(0)$ are therefore Gaussian distributed \cite{Jain2016}. This means that every iteration of the free-fall cycle results in different oscillation amplitudes after the free fall. In order to estimate the expectation value of the COM energy, we average the measured COM energies after the free fall over 1000 iterations. We plot the mean COM energy for the horizontal $x$ oscillation for varying fall durations between $\tau=10\,\mathrm{\mu s}$ and $140\,\mathrm{\mu s}$ in Fig.~\ref{fig:energies_free_fall}(d) as black dots. We find that the mean $x$ oscillation energy scales quadratically with the fall duration $\tau$ (dotted lines), which is expected as the initial velocity results in a displacement which is linear in time. In contrast, for the mean oscillation energy along the vertical $y$ axis, shown in Fig.~\ref{fig:energies_free_fall}(e), we observe a significant deviation from this quadratic scaling for long fall durations, which originates from the acceleration of the particle due to gravity.

\begin{figure} 
	\includegraphics{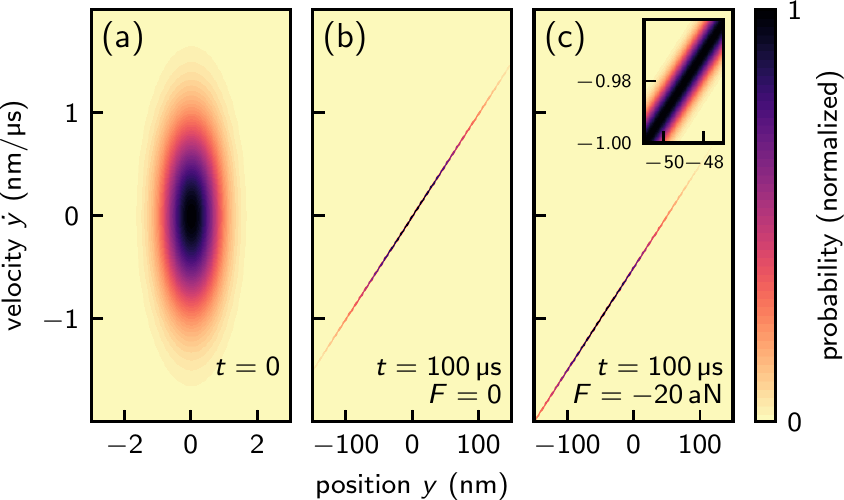}
	\caption{(a)~Initial probability distribution in phase space, corresponding to time $t=0$ in Fig.~\ref{fig:energies_free_fall}(c), shown for a thermal state with a COM energy of $E_0/k_\mathrm{B}=50\,\mathrm{mK}$. (b)~Directly after a free evolution of duration $\tau=100\,\mathrm{\mu s}$ with deactivated trapping potential and in the absence of any force, the distribution is spread along the position axis due to the initial velocity of the particle. (c)~Phase-space distribution after the evolution of duration $\tau=100\,\mathrm{\mu s}$ in the presence of a static force $F=-20\,\mathrm{aN}$. Compared to (b), the distribution is displaced by $\Delta y=F\tau^2/(2m)=-49\,\mathrm{nm}$ and $\Delta \dot{y}=F\tau/m=-0.98\,\mathrm{nm/\mu s}$ (inset). \label{fig:phase-space}}
\end{figure}

To gain a quantitative understanding of the mean oscillation energy after the free fall, we study the evolution of the particle's phase-space distribution during the fall. The following derivation is given for the $y$ axis, but remains equivalently valid for the other oscillation directions. For $t<0$, the particle's COM motion is in a thermal state with energy $E_0$. In position-velocity phase-space, just before the free fall, this results in a Gaussian probability distribution 
\begin{equation}\label{eq:phase-space-gaussian}
\mathcal{P}(y, \dot{y};t=0)=\frac{1}{Z}\exp\left[-\frac{m\Omega_0^2y^2+m\dot{y}^2}{2E_0}\right]
\end{equation}
that is centered at $y=0$ and $\dot{y}=0$ with $Z=2\pi E_0/(m\Omega_0)$ \cite{Jain2016}. In Fig.~\ref{fig:phase-space}(a), we plot this phase-space distribution for an initial energy of $E_0/k_\mathrm{B}=50\,\mathrm{mK}$ and an oscillation frequency of $\Omega_0=2\pi\times 160\,\mathrm{kHz}$. 

During the free fall of duration $\tau$, the particle's COM is accelerated due to the static force $F$. The position at time $t=\tau$ becomes $y(\tau)=y(0)+\dot{y}(0)\tau+F\tau^2/(2m)$ and the velocity reads $\dot{y}(\tau)=\dot{y}(0)+F\tau/m$. In the absence of a force ($F=0$), propagating Eq.~(\ref{eq:phase-space-gaussian}) results in the phase-space distribution displayed in Fig.~\ref{fig:phase-space}(b). The probability distribution is not Gaussian anymore and spreads towards larger positions, because high initial velocities translate to large displacements after the free fall. In the presence of a force $F=-20\,\mathrm{aN}$ acting on the particle, the phase-space distribution from Fig.~\ref{fig:phase-space}(b) is additionally shifted towards negative positions and negative velocities due to the acceleration the particle experiences during the free fall [see Fig.~\ref{fig:phase-space}(c)]. By integrating $\iint\mathcal{P}(y, \dot{y}; \tau)E(y, \dot{y})\,\mathrm{d} y\,\mathrm{d}\dot{y}$ over the entire phase-space, we derive the expectation value of the oscillation energy after the free fall \cite{Ross2014}
\begin{eqnarray}
\langle E \rangle (\tau)
&=&E_{0}+\frac{E_{0}\Omega_{0}^2}{2}\tau^2 + \frac{F^2}{2m} \tau^2 +\frac{F^2\Omega_{0}^2}{8m}\tau^4.\label{eq:expvalue_energy}
\end{eqnarray}
The first term on the right-hand side of Eq.~(\ref{eq:expvalue_energy}) is the initial energy of the COM motion, the second term is the potential energy originating from the displacement due to the initial velocity of the particle, and the two last terms are the kinetic and potential energy the particle acquires due to the force that accelerated its motion. Hence, a free evolution, in the absence of a force, results in a mean energy that scales quadratically with the fall duration $\tau$. In contrast, if a static force $F$ is present, the mean oscillation energy scales with $\tau^4$ for long fall durations.

We fit $\bar{E}=\beta_0+\beta_2\tau^2+\beta_4\tau^4$ to the measured mean oscillation energies in Fig.~\ref{fig:energies_free_fall}(d) for the $x$ and in Fig.~\ref{fig:energies_free_fall}(e) for the $y$ axis (solid lines). Using the fit parameter $\beta_2$, we first derive the mean initial oscillation energy $E_0=2\beta_2/\Omega_0^2$, and find $E_{x,0}/k_\mathrm{B}=37(2)\,\mathrm{mK}$ for the $x$ axis and $E_{y,0}/k_\mathrm{B}=37(2)\,\mathrm{mK}$ for the $y$ axis. Second, we deduce the force that acts on the particle during the free fall $|F_y|=\sqrt{8m\beta_4/\Omega_0^2}=21(4)\,\mathrm{aN}$, and derive a gravitational acceleration of $g=10.4(18)\,\mathrm{m/s^2}$, which is in agreement with the textbook value $g=9.8\,\mathrm{m/s^2}$.


\paragraph{Electrostatic force measurement.} 

\begin{figure}[b] 
	\includegraphics{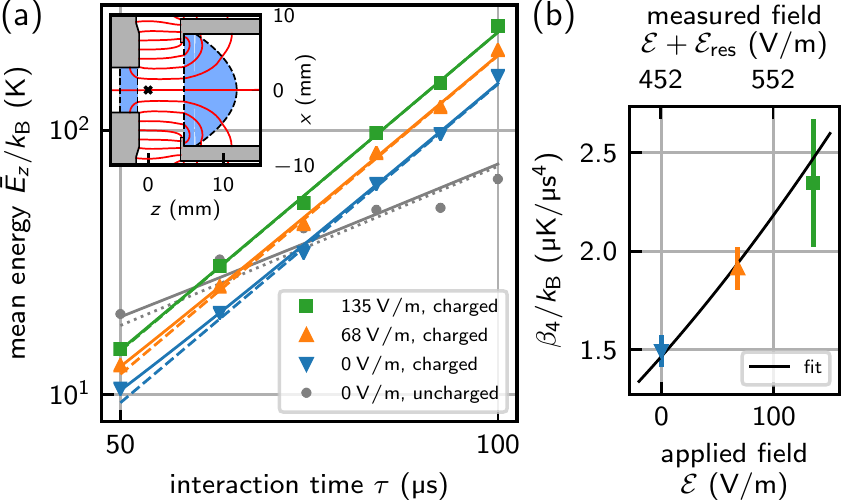}
	\caption{(a) Mean energy in $z$ direction as a function of interaction time $\tau$ at different electric field strengths (squares and triangles). We fit $\bar{E}=\beta_0+\beta_2\tau^2+\beta_4\tau^4$ (solid lines) and indicate the $\beta_4\tau^4$ component of the fit (dashed) for the data taken with a charged particle. The energy of an uncharged particle (gray points) scales with $\tau^2$ (dotted line, the initial energy is higher than in the charged case).
Inset: Sketch of capacitor formed by objective and collection lens holder (metal gray, glass blue), and field lines (red).
	(b) Fitting parameters $\beta_4$ from (a) with $\beta_4(\mathcal{E})=q^2(\mathcal{E}+\mathcal{E}_\mathrm{res})^2\Omega_0^2/(8m)$ (black). The deduced fitting parameter is the residual electric field strength $\mathcal{E}_\mathrm{res}=452(13)\,\mathrm{V/m}$. \label{fig:energies_charge}}
\end{figure}

Having demonstrated the ability to detect the gravitational force acting on the nanoparticle, we apply our static-force sensing technique to measure a Coulomb force. To this end, we adjust the net charge on the nanoparticle to a single elementary charge \cite{Moore2014,Frimmer2017}. By applying a voltage of $1\,\mathrm{V}$ to a capacitor formed by the objective and the holder of the collection lens, we generate an electric field along the $z$ direction \cite{Ranjit2016}. According to a finite element simulation, the field points along the $z$ direction and the field strength at the position of the particle is $\mathcal{E}=135\,\mathrm{V/m}$ [see inset Fig.~\ref{fig:energies_charge}(a)], which corresponds to an expected electrostatic force of $F=22\,\mathrm{aN}$ \cite{Frimmer2017}. In Fig.~\ref{fig:energies_charge}(a), we plot the average oscillation energy in $z$ direction after the interaction for 1000 free-fall cycles and for varying interaction times $\tau$ (green squares). We find a clear $\tau^4$ scaling for interaction times exceeding $50\,\mathrm{\mu s}$, which indicates the presence of a force along the $z$ axis. When we reduce the applied electric field to $\mathcal{E}=68\,\mathrm{V/m}$ (orange triangles), and to $\mathcal{E}=0\,\mathrm{V/m}$ (blue triangles), the measured oscillation energy reduces. Surprisingly, even when the capacitor field is switched off ($\mathcal{E}=0\,\mathrm{V/m}$), we still observe a clear $\tau^4$ dependence of the mean oscillation energy, meaning that there is a residual field accelerating the charged particle. In contrast, when using an uncharged particle and turning off the capacitor field, the mean oscillation energy scales as $\tau^2$ (gray dots), which means, that no force accelerates the particle in this case. The residual field that we measure with a charged particle originates from 
contact (Volta) potentials \cite{Bagotsky2005} and stray fields due due patch charges on close-by surfaces \cite{Brownnutt2015}. 
%
To estimate the strength of the residual field $\mathcal{E}_\mathrm{res}$, we plot the fitting parameter $\beta_4$ for the three field amplitudes in Fig.~\ref{fig:energies_charge}(b). By fitting a quadratic function (black) to the data points, we find a residual electric field of $\mathcal{E}_\mathrm{res}=452(13)\,\mathrm{V/m}$ when the capacitor field is switched off. This residual field corresponds to a residual force of $F_\mathrm{res}=72(2)\,\mathrm{aN}$. For the case, where we apply a field of $\mathcal{E}=135\,\mathrm{V/m}$ (blue), we deduce a force of $F=\sqrt{8m\beta_4/\Omega_0^2}=92(17)\,\mathrm{aN}$, which means that we measure an additional force acting on the particle of $20(7)\,\mathrm{aN}$ when applying the capacitor field, which agrees with the expected value.

\paragraph{Discussion.} 
The sensitivity of the static-force measurement scheme demonstrated in this work is limited by the initial energy $E_0$ of the particle when the optical trap is deactivated, and by the interaction time $\tau$. To detect a static force $F$, we require the fourth term in Eq.~(\ref{eq:expvalue_energy}) to exceed the second term, \emph{i.e.}, $F>\sqrt{4E_0m}/\tau$. Therefore, we benefit from the small mass of the nanoparticle and from our ability to efficiently cool the particle's COM energy \cite{Li2011,Gieseler2012}. As previously shown, further improvements in the cooling performance to reach initial oscillation energies of $E_0/k_\mathrm{B}<1\,\mathrm{mK}$ are feasible \cite{Jain2016}. Even cooling to the quantum ground state of motion seems within reach. For an initial energy of $1\,\mathrm{mK}$, we estimate that the interaction time is limited to $\sim 300\,\mathrm{\mu s}$ by our ability to retrap the particle along the fall direction after the free fall, which would make the detection of static forces as low as $1\,\mathrm{aN}$ possible. For even longer interaction times we envision the use of a double-trap configuration for retrapping the particle after the free fall \cite{Rondin2017}. Alternatively, a compensation of the gravitational force is possible using an electrostatic force generated by a suitable electrode configuration. For a ground-state cooled particle, we therefore estimate an ultimate limit of the interaction time of $\tau=100\,\mathrm{ms}$, i.e., a detection limit for static forces of less than $1\,\mathrm{zN}$.

\paragraph{Conclusion.} 
We have demonstrated the measurement of a $10\,\mathrm{aN}$ static force using a levitated nanoparticle. Our technique is based on a free interaction of the particle with the static force, while the trapping potential is deactivated, and a resonant read-out of the subsequent oscillation. 
Because the particle's dynamics is measured along three orthogonal axes the scheme is applicable to the mapping of vectorial force fields.
In contrast to comparable experiments using clouds of cold atoms \cite{Baumgartner2011}, we are able to perform our measurements repeatedly with a single particle, and measure the position with higher precision. Furthermore, the high mass density of the levitated silica spheres paves the way to the investigation of short-range forces, such as Casimir or van-der-Waals forces with an uncharged particle close to a surface \cite{Geraci2010}. Finally, the demonstration of a controlled free-fall experiment also marks an important step towards the realization of quantum-interference experiments and time-of-flight state tomography, which are essential building blocks for realizing non-Gaussian quantum states in macroscopic objects \cite{Romero-Isart2011,Wan2016}.

\begin{acknowledgments}
This work has been supported by  ERC-QMES (No. 338763) and the Swiss National Centre
of Competence in Research (NCCR) -- Quantum Science and Technology (QSIT) program (No. 51NF40-160591).
\end{acknowledgments}

%

\end{document}